\documentclass[page-classic]{epl2} 

\usepackage{amsmath}
\usepackage{color}
\usepackage{ulem}

\newcommand{\marge}[1]{\marginpar{}}  % do not show margin notes
\newcommand{\Sl}[1]{{}}           % do not show labels
\newcommand{\beq}[1]{\Sl{#1}\begin{equation}\if#1\empty\else\label{#1}\fi}
\newcommand{\eeq}{\end{equation}}
\newcommand{\beqa}[1]{\Sl{#1}\begin{eqnarray}\if#1\empty\else\label{#1}\fi}
\newcommand{\eeqa}{\end{eqnarray}}

\shorttitle{Nonextensive statistics and continuous Hamiltonian systems?}
\institute{
  \inst{1} Physics Department,  CP 231,
  Universit\'{e} Libre de Bruxelles, 1050 Brussels, Belgium 
}

\title{Comment on ``Possible divergences in Tsallis' thermostatistics''.}
\author{ James F. Lutsko$^1$\thanks{E-mail: \email{jlutsko@ulb.ac.be}}
and Jean Pierre Boon$^1$\thanks{E-mail: \email{jpboon@ulb.ac.be}}}
\shortauthor{ J.F. Lutsko and J.P. Boon}


\abstract{In a recent letter ({\it{EPL}}, {\bf{104}} (2013) 60003; see also {\it {arXiv:1309.5645}}), Plastino and 
Rocca suggest that the divergences inherent to the formulation of nonextensive statistical mechanics
can be eliminated {\it {via}} the use of  $q$-Laplace transformation which is illustrated for the
case of a kinetic Hamiltonian system, the harmonic oscillator. The suggested new formulation
raises questions which are discussed in the present comment.}

\begin{document}

\maketitle

The nonextensive statistical mechanics introduced by Tsallis\cite{Tsallis88} and developed
over the last 25 years by numerous researchers \cite{Tsallis2009} is based on a generalization of
the Boltzmann entropy,%
\begin{equation}
S_q=\frac{1}{q-1}\left(  1-\sum_{n}p_{n}^{q}\right)  \rightarrow\frac{1}%
{q-1}\left(  1-\int p^{q}\left(  x\right)  dx\right) %
\underset{q\rightarrow1}{\Rightarrow} -\int p\left(  x\right) \log \,p\left(  x\right)  dx \,,
\end{equation}
where we have given the expressions for both discrete and continuous
variables. Statistical mechanics is then developed using the maximum entropy
formalism whereby the probabilities are determined by maximizing the entropy
subject to the constraint of constant average energy and normalizability of the
distribution function. The result is a so-called $q$-exponential distribution%
\begin{equation}
p_{n}=Z_{q}^{-1} \exp_{q}\left(\beta \left(\varepsilon_{n}-U\right)\right) 
\equiv Z_{q}^{-1}\left(  1-\left(  1-q\right)  Z_{q}^{q-1}\beta\left(
\varepsilon_{n}-U\right)  \right)  _{+}^{\frac{1}{1-q}}  \,,%
\end{equation}
where $Z_{q}$ is determined by normalization, $\beta$ is the inverse
temperature, $\varepsilon_{n}$ is the energy of state $n$, $U$ is the average
total energy and the notation $\left(  x\right)  _{+}^{y}$ means $x^{y}$ when
$x>0$ and zero otherwise. The expression for the continuous case is analogous. 
In both cases, the distribution becomes the usual exponential, Maxwell-Boltzmann 
distribution in the limit $q \rightarrow 1$.
Lutsko and Boon noted that for Hamiltonian systems \cite{LutskoBoon}, the continuous
distribution is only normalizable for values of the parameter $q$ satisfying
$0\leq q \leq 1+ \frac{2}{N D}  $ where $N$ is the number of particles
and $D$ the dimension of the system.
Since it is the case that $q>1$ corresponds to so-called fat-tailed
distributions observed in many physical and non-physical systems (see Part III in
\cite{Tsallis2009}) and so is of most interest, this places a significant constraint
on the utility of the formalism for many-body systems ($N>>1$). 
The divergence of the normalization results
from the combination of the power-law distribution (2) and the unbounded nature of
the kinetic energy \cite{LutskoBoon}.

Recently, Plastino and Rocca \cite{PlastinoRocca} have proposed a modification of 
the Tsallis formalism that is intended to circumvent this problem. They note that in the
usual, Boltzmann-Gibbs, statistical mechanics, the normalization factor, or partition function\footnote{We note that Plastino and Rocca (PR) identify the normalization factor with the partition function without comment. Although it is not our main point, this cannot be correct since the normalization factor has dimensions and the partition function should be dimensionless. In the usual formulation, there is an additional factor of $\frac{1}{h^{DN}N!}$ relating these quantities  where $h$ is Planck's constant. We will follow PR in ignoring this distinction.}, can be written
in the form
\begin{equation}
Z_{q=1}=\int_{0}^{\infty}e^{-\beta U}g\left(  U\right)  dU,\;\;g\left(  U\right)
\equiv\int\delta\left(  U-H\left(  \Gamma\right)  \right)  d\Gamma \,,
\end{equation}
where $H\left(  \Gamma\right)  $ is the Hamiltonian (assumed to be shifted so
as to be bounded below by zero) and $\Gamma$ represents a point in phase
space. So, the partition function can be viewed as a Laplace transform of the
density of states. Similarly, the average of any function of the Hamiltonian,
$\left\langle f\left(  H\right)  \right\rangle $, can be written in similar
form with the replacement of $g\left(  U\right)  $ by $f\left(  U\right)
g\left(  U\right)  $: in particular, this applies to the average energy and to
the entropy. What is proposed in \cite{PlastinoRocca} is to eliminate the divergences 
by replacing the Laplace transform  structure by the so-called $q$-Laplace transform 
\cite{RoccaPlastino} defined for an arbitrary function $f(x)$ as
\begin{equation}
\widetilde{f}_{q}\left(  \alpha\right)   
 \equiv\Theta\left(  \operatorname{Re}%
\left(  \alpha\right)  \right)  \sum_{n}a_{n}\int_{0}^{\infty}x^{n}\left(
1-\left(  1-q\right)  \alpha x^{n\left(  q-1\right)  }\right)  ^{\frac{1}%
{1-q}}dx\label{f} \,,%
\end{equation}
where it is assumed that the function $f(x)$ can be expanded about $x=0$ with
coefficients $a_{n}$ (In fact, the expression given above is the sum of the
$q$-Laplace transforms of individual terms $x^{n}$ but this distinction is not
important). Hence, the proposed partition function is $Z_{q} = \widetilde{g}_q(\beta)$, 
with similar expressions for the energy and entropy. It is then shown  that for a 
harmonic oscillator these expressions are all finite.

Careful examination of the procedure developed in \cite{PlastinoRocca} raises
%We believe that there are several problems with this procedure 
several problems which makes it questionable as a basis for statistical mechanics:

\begin{enumerate}
\item While this procedure  yields a finite value for the partition function
$Z_q$ (Eq.(23) in \cite{PlastinoRocca}), the original distribution 
$f_q \sim \exp_q(-\beta (H-U))$ remains un-normalizable (beyond the domain 
$0\leq q \leq 1+\frac{2}{N D}  $). It therefore fails to address
the fundamental problem with the nonextensive Tsallis formalism. Introducing the 
$q$-Laplace prescription only ``cures'' averages of functions of the energy.

\item If the "partition function" is not related to the normalization of the
distribution, we assume its finiteness is only important because it is related
to the free energy in the usual way. Indeed, in this modified formalism, one
still finds that $U-TS=-k_{B}T\ln Z_{q}$ so that this should be identified as the
free energy. However, in this case  the modified free energy  and internal
energy do not satisfy the thermodynamic relation  $U=-\left(  \partial
F/\partial\beta\right)  _{V}$. 

\item It is stated in the third section  of \cite{PlastinoRocca} that 
in the nonextensive approach the corresponding values
for the partition function, the mean energy, and the entropy 
can be obtained by replacing the quantities appearing  in the classical statistical 
thermodynamics expressions by their $q$-analogues.\footnote{Note however that
the expression given for the  entropy, Eq.(11) of Ref.\cite{PlastinoRocca}, 
differs from the Tsallis definition where the first factor of $P$ is raised to the power $q$.} 
Now the resulting ``entropy'' is not equivalent to the original Tsallis  entropy
evaluated with the $q$-exponential distribution (compare Eqs.(15) and (21) of Ref.\cite{PlastinoRocca}). Nor is it an alternative to the Tsallis entropy since it is not a 
{\it{functional}} of the distribution but in fact only applies to the distribution derived 
from maximization of the Tsallis entropy. What is the justification for using the Tsallis 
entropy to determine the distribution but another ``entropy'' to define the thermodynamics? 

\item The expansion used in Eq.(\ref{f}) above seems quite arbitrary. One
could, for example, replace $a_{n}x^{n}$ by $\left(  2^n a_{n}\right)  \left(
\frac{x}{2}\right)  ^{n}$ and thereby obtain an inequivalent form for the
function $\widetilde{f}_{q}\left(  \alpha\right)  $. In fact, this problem is
evident in the proposed thermodynamic expressions since they involve
quantities of the form $\beta U^{n\left(  q-1\right)  +1}$ which should be
dimensionless but are not. One can "solve"\ this problem by replacing
$a_{n}U^{n}$ by $b_{n}\left(  \frac{U}{u}\right)  ^{n}$ with $u$ a constant
having the dimensions of energy and $b_{n}\equiv u^{-n}a$ but the results then
depend on the choice of $u$. 
\end{enumerate}

In conclusion, we have analyzed the proposal of a modified formulation of
nonextensive statistical thermodynamics based on the use of the $q$-Laplace
transform in order to eliminate divergences related to the non-normalizability of
the $q$-distribution function introduced in \cite{Tsallis88}.
While the modifications proposed in \cite{PlastinoRocca} do indeed produce finite 
quantities,  the fundamental problem of the divergence of the normalization of the 
$q$-distribution function remains unsolved and it is unclear whether the new 
formulation can produce a consistent thermodynamics.

\acknowledgments
This work of JFL was supported by the European Space Agency under contract
number ESA AO-2004-070. 
%\end{acknowledgments}

\bigskip

%\bibliographystyle{eplbib}
%\bibliography{unstable}

\end{document}